\documentclass[english,aps,epsfig,graphics,twocolumn,floatfix,mathbbm,a4paper,nofootinbib]{revtex4}
\usepackage[T1]{fontenc}
\usepackage[latin9]{inputenc}
\usepackage{amsmath}
\usepackage{amssymb}
\usepackage{color}
\makeatletter
\@ifundefined{textcolor}{}
{%
 \definecolor{BLACK}{gray}{0}
 \definecolor{WHITE}{gray}{1}
 \definecolor{RED}{rgb}{1,0,0}
 \definecolor{GREEN}{rgb}{0,1,0}
 \definecolor{BLUE}{rgb}{0,0,1}
 \definecolor{CYAN}{cmyk}{1,0,0,0}
 \definecolor{MAGENTA}{cmyk}{0,1,0,0}
 \definecolor{YELLOW}{cmyk}{0,0,1,0}
}

\newtheorem{theorem}{Theorem}
\newtheorem{corollary}{Corollary}
\newtheorem{proposition}{Proposition}

\newtheorem{lem}{Lemma}

\newcommand{\un}{1\mkern -4mu{\rm l}}

\usepackage{babel}

\makeatother

\usepackage{babel}
\begin{document}

\title{Unbounded violation of quantum  steering inequalities }

\author{M. Marciniak$^{1}$, A. Rutkowski$^{1,2}$, Z. Yin$^{3,1}$,  M. Horodecki$^{1,2}$ and R. Horodecki$^{1,2}$}
\affiliation{
$^1$ Institute of Theoretical Physics and Astrophysics, University of Gda\'nsk, 80-952 Gda\'nsk, Poland \\
$^2$ National Quantum Information Centre of Gda\'nsk, 81-824 Sopot, Poland\\
$^3$ School of Mathematics and Statistics, Wuhan University, China
}

\begin{abstract}
We construct steering inequalities which exhibit unbounded violation. The concept was to exploit the relationship between steering violation and uncertainty relation. To this end we apply mutually unbiased bases  and anti-commuting observables, known to exibit the strongest uncertainty. In both cases, we are able to procure unbounded violations. Our approach is much more constructive and transparent than the operator space theory approach employed to obtain large violation of Bell inequalities. Importantly, using anti-commuting observables  we are able to obtain a {\it dichotomic} steering inequality with unbounded violation. So far  there is no  analogous result for  Bell inequalities. Interestingly,  both the dichotomic inequality  and  one of our inequalities can not be directly obtained from existing uncertainty relations, which strongly suggest the existence of an unknown kind of uncertainty relation.

\end{abstract}


\maketitle

\emph{Introduction}. Quantum theory is the primary mainstay of our understanding and formal description of nature. Moreover, it constitutes a perfect empirically confirmed formal construction.
Despite many years of continuous attempts, a commonly accepted interpretation of mathematical formalism of Quantum Mechanics has not been found. The phenomenon of quantum correlations, especially entanglement, is believed
to be most amazing and eluding the schemes of classical thinking.
Multiannual conceptual efforts to grapple the 'spooky actions for spatially separated  systems' began with the fundamental work of Einsten, Podolski
and Rosen \cite{EPR} and continue until this day. Nowadays, we possess the knowledge that quantum correlations - still remaining a great mystery- allow experimental realization.  Additionally, they can be controlled and implemented in nontrivial tasks. Secure quantum communication as well as quantum calculations are amongst them. Such promising perspectives to practically use quantum correlations as
a resource, clearly demonstrate the importance of the undertaken efforts to
improve our deep understanding of this phenomenon.

The concept of \emph{quantum steering} was first introduced by Schr\"{o}dinger
in 1935 \cite{schrod} as a generalization of EPR paradox \cite{EPR}
for bipartite systems in arbitrary pure entangled states and arbitrary
measurements by one party. Consider two separated observers sharing entanglement.
The first observer, by  measurement on his system, can steer the state of the system held by the second
observer. 
Like the debate of EPR paradox, the notion of quantum steering had been ignored for a long time until it was recovered by H. M. Wiseman, S. J. Jones, and A. C. Doherty \cite{wieseman2007}, where they introduced quantum steering as an information task. Like in the Bell scenario,
the non-classicality revealed by the steering phenomenon is expressed by means of  violation of the so called
{\it steering inequalities}.
It should be noticed that not all entangled states lead to steering, and  there are states which violate steering inequalities,
but do not violate any  Bell inequality  \cite{wieseman2007, QVCADAB2015}.

Recently, {\it unbounded} violations of Bell inequalities were intensively analysed, mostly by means of advanced tools of mathematical physics  \cite{JPPVW2010,JP2010prl,JP2011} as well as communication complexity methods \cite{briet,buhram2012}.
The existing results are either mostly random constructions having origin in the existing knowledge from the field of operator spaces,
or are derivatives of quite complicated communication complexity protocols.

In this paper we analyse the unbounded violation of steering inequalities. We exploit an intrinsic relationship of steering phenomenon with the uncertainty principle (see e.g. \cite{up2015}), and apply the measurements that offer strong uncertainty, such as mutually unbiased bases (MUBs) and Clifford observables
\cite{WW}. Until now it was not known if quantum correlations type steering are equivalent to Bell type correlations in the regime of large violation. Here, we provide two results that address this issue:
(i) using mutually unbiased basis we obtain larger violation than the largest quantum violation of Bell inequalities,
(ii) by means of  Clifford observables we  provide unbounded violation of steering inequality with binary outputs - a feature which is still unknown for Bell inequalities with binary outputs for one of the parties.

Our inequalities are extremely simple in comparison to the existing Bell inequalities exhibiting large violation  \cite{JP2011,JPPVW2010,Regev2012} as well as to random constructions of steering inequalities based on the operator space approach, provided in the companion paper  \cite{HMY2014}.
While one of our violations is a consequence of existing fine-grained uncertainty principle \cite{OW2010}
obtained in \cite{R2014}, our other results -- the unbounded violation for binary observables and a variant of large violation with MUBs -- cannot be derived from any existing uncertainty principles.

\emph{Steering inequality}. Hereunder we will consider the following steering scenario \cite{Pusey2013,HMY2014}, which is equivalent to the one in \cite{wieseman2007}. Suppose there are two observers (Alice and Bob). Alice can choose among $n$ different measurement settings labeled by $x=1,\ldots,n.$ Each of which can result in one of $m$ outcomes, labeled by $a =1, \ldots, m.$ Suppose the local Hilbert space dimension for Bob is $d.$  The available data are the steered states, they are positive operators on $\mathcal{H}_{B}:$   $\sigma_{x}^{a}\geq0$ , and by the no-signaling principle \cite{B2005} we have that $\text{Tr}\left(\sum_{a}\sigma_{x}^{a}\right)=1$ and it is independent
of $x$. We will denote the set of those operators as $\sigma=\{\sigma_{x}^{a}:\, x=1,\ldots,n,\; a=1,\ldots,m\}$ and call it $\left( n,\, m,\,\dim\left(\mathcal{H}_{B}\right)\right)$ - assemblage or simply assemblage. The set of all assemblages will be denoted by $\mathcal Q$. It is well known \cite{Sch1936,HJW1993} that any assemblage $\sigma$ has a {\it quantum realization}, i.e. it can be generated remotely, by performing measurements on a subsystem of  bipartite quantum states. More precisely,
for any assemblage $\sigma$ there exists a Hilbert space $\mathcal{H}_{A}$ such
that
\begin{equation}\label{eq:qassem}
\sigma_{x}^{a}=\text{Tr}_{A}((E_{x}^{a}\otimes\un_{B})\rho),
\end{equation}
for every $x$ and $a$, where $\rho\in B(\mathcal{H}_{A}\otimes\mathcal{H}_{B})$
is a density matrix and $\{E_{x}^{a}\}_{a=1}^{m}\subset B(\mathcal{H}_{A})$
( by $B(\mathcal{H})$ we mean the algebra of all bounded linear operators
on $\mathcal{H}$ ) is a POVM measurement on Alice for every $x$,
i.e. $E_{x}^{a}\geq0$ for every $x,a$, and $\sum_{a}E_{x}^{a}=\un,$
for every $x$.

If the shared state is separable, by measuring its subsystems, one can only generate assemblages that possess local hidden state model, defined as follows.
The assemblage has a \textit{local
hidden state (LHS)} model, if there is a finite set of indices $\Lambda$,
nonnegative coefficients $q_{\lambda}$ such that $\sum_{\lambda}q_{\lambda}=1$,
density matrices $\sigma_{\lambda}$ in $B(\mathcal{H}_{B})$ for
$\lambda\in\Lambda$, and probability distributions $\{p_{\lambda}(a|x)\}_{a}$
for every $x$ and $\lambda$ (i.e. $p_{\lambda}(a|x)\geq0$ and $\sum_{a}p_{\lambda}(a|x)=1$
for every $x,\lambda$), such that
\begin{equation}\label{eq:LHS}
\sigma_{x}^{a}=\sum_{\lambda\in\Lambda}q_{\lambda}p_{\lambda}(a|x)\sigma_{\lambda},
\end{equation}
for every $x,a$. We denote the set of LHS assemblages by $\mathcal{L}.$

As a Bell functional (inequality) can be used to show the incompatibilies between the local hidden variable (LHV) model and the quantum theory, we can use the steering inequalities \cite{Cavalcanti} to study the difference between the two sets $\mathcal{L}$ and $\mathcal{Q}.$ 
 Firstly, let us define a steering inequality in the spirit of \cite{Pusey2013}. Let $F$ be some function from $\mathcal Q$-assemblages to the real numbers. If $S_{LHS}(F)$ is the maximum of $S$ over all assemblages that admit LHS models then $S\leq S_{LHS}(F)$ is called a \textit{steering inequality}. Let $S_{\mathcal Q}(F)$ be the maximum of $F$ over all assemblages (recall that all assemblages have a quantum realization \cite{Sch1936,HJW1993}). If $S_{\mathcal Q}(F)> S_{LHS}(F)$ then the steering inequality is called nontrivial, i.e. it can be violated  using entangled states.  We will consider only the linear functional from the space of assemblages to the real numbers. In other words, we can define the steering functional  in the following way: for  given natural numbers $n,\, m$ and $d$, we define a
steering functional  $F$ as a set $\left\{ F_{x}^{a}:\, x=1,\ldots,n,\; a=1,\ldots,m\right\} $
of $d\times d$ real matrices. For a given assemblage $\sigma$
we get a real number
\begin{equation}\label{eq:steeringineq}
\left\langle F,\sigma\right\rangle  =\mathrm{Tr}\left(\sum_{x=1}^{n}\sum_{a=1}^{m}F_{x}^{a}\sigma_{x}^{a}\right).
\end{equation}

Additionally, let us define two quantities: for a given steering functional $F$, we define
the LHS bound of $F$ as the number

\begin{equation}
S_{LHS}(F)=\sup\left\{ \left|\left\langle F,\sigma\right\rangle\right|: \; \sigma\in\mathcal{L}\right\} ,
\end{equation}
and the quantum bound of $F$ as

\begin{equation}
S_{Q}(F)=\sup\left\{ \left|\left\langle F,\sigma\right\rangle\right|: \; \sigma\in\mathcal{Q}\right\}.
\end{equation}
Now we are ready to define the quantum violation of $F$ as
the number

\begin{equation}
V\left(F\right)=\frac{S_{Q}\left(F\right)}{S_{LHS}\left(F\right)}.
\end{equation}

A steering functional with large violation, will tell us the sets $\mathcal{L}$ and $\mathcal{Q}$ are prominently different. Apart from the above theoretical aspect, there will be many benefits when we apply it to practical experiments \cite{SJWP2010, Smith2012} and applications \cite{BCWSW2012}. However, for a given Bell or steering functional, it is difficult to calculate its violation. Operator space theory was shown to be a powerful tool to overcome this difficulty. See \cite{JPPVW2010,JP2011} in Bell scenario and \cite{HMY2014} in steering. For example, in scenario $(d,d,d),$ the following random steering functional was considered in the campanion article \cite{HMY2014}:
\begin{equation}
F_x^a = \frac{1}{d} \sum_{k=1}^d \epsilon_{x,a}^k \left| 1 \left\rangle \right\langle k \right|, \;\; x,a=1,\cdots,d,
\end{equation}
where $\epsilon_{x,a}^k, x,a,k=1,\cdots,d$ are independent Bernoulli variables. The violation of this inequality is $O (\sqrt{\frac{d}{\log d}})$.

In this paper, we are able to derive steering functionals by using MUBs and Clifford algebra,
More precisely, for the scenario $(d+1, d, d),$ when dimension of Hilbert space $d$ is equal power of prime number then we know there exits exactly $d + 1$ MUBs) by using MUBs, we can construct a steering functional with unbounded violation of order $O(\sqrt{d}).$ It can be seen that we can obtain larger violation compared to the random one. On the other hand, for the scenario $(n,2,2^n),$ we are able to find a \emph{dichotomic} steering functional with unbounded violation of order $O\sqrt{\frac{n}{2}}) \simeq O\left(\sqrt{\log d}\right)$  by using Clifford observables. Therefore, this unbounded violation reveals an interesting and particular property of quantum steering.

\emph{Unbounded violation: Mutually unbiased bases}. Now we are going to study a steering functional constructed by means of mutually unbiased bases (MUBs)
(\cite{WF1989}).
Let $M_{1}$ $=\left\{ \left|\phi_{1}^{a}\right\rangle :\, a=1,\ldots, d\right\} $
and $M_{2}$ $=\left\{ \left|\phi_{2}^{a}\right\rangle :\, a=1,\ldots, d\right\} $
be orthonormal bases in the $d-$dimensional Hilbert space. Then they
are said to be mutually unbiased if $\left|\left\langle \phi_{1}^a|\phi_{2}^b\right\rangle \right|=\frac{1}{\sqrt{d}}$
for all $a,\, b=1,\ldots,d.$ A set $M=\left\{ M_{x}:\, x=1,\ldots n\right\} $
of orthonormal bases of $\mathbb{C}^{d}$ is said to be a set of mutually
unbiased bases (MUBs), if $M_{x}$ and $M_{y}$ are mutually unbiased for every $x\neq y$.


Given MUBs $M,$ we define the steering functional $F=\{F_x^a\}$, where
\begin{equation}
\label{eq:steerineqMUBs}
F_x^a=|\phi_x^a\rangle\langle\phi_x^a|,\qquad x=1,\ldots,n,\;a=1,\ldots,d.
\end{equation}
Our aim is to calculate $V(F)$. Firstly, we would like to estimate the quantity $S_Q(F)$. We propose the following
\begin{lem}
\label{qvalue}
Let $F$ be the steering inequality defined in \eqref{eq:steerineqMUBs}. Then
\begin{eqnarray}
S_{Q}\left(F\right) = 
n,\label{eq:Bquantum}
\end{eqnarray}
and the  maximal value is attained on maximally entangled state.
\end{lem}
Proof of this lemma is in appendix A. To estimate the bound of $S_{LHS}$, we use the   operator norm estimation  of some operator (see appendix A).   Compare this result with  Lemma \ref{qvalue} we are ready to formulate one of the main results.

\begin{theorem}
\label{thm:MUB} If $F$ is a steering functional determined by MUBs as in \eqref{eq:steerineqMUBs},  then we have

\begin{equation}
V\left(F\right)
\geq \frac{n\sqrt{d}}{n+1+\sqrt{d}}.
\end{equation}
\end{theorem}
 Proof of this theorem is in appendix A. If the dimension $d$ is an integer power of a prime number,  then we can  always find $d+1$ MUBs \cite{WF1989}. In this case $n=d+1;$ hence, we can find a steering functional $F,$ with violation $O(\sqrt{d}).$ It is better than the random one in the sense that it has a higher order of violation.

Our result is connected to the results obtained in \cite{OW2010}. In 
that paper the authors revealed that "nonlocality of quantum mechanics and Heisenberg's uncertainty principle are inextricably and quantitatively linked." They introduced a notion named "fine-grained uncertainty relations" to characterise the "amount of uncertainty" in a particular physical theory. For the given set of measurements $A_{x}$, with $ x=1,\ldots ,n$ and the set of outputs $\vec{a}=\left\{ a(x):\, x=1,\ldots,n\right\} $, consider the following quantity introduced in \cite{OW2010} 
\begin{equation}\label{eq:uncert}
 \xi_{\vec{a}}=\max_{\rho} \left\{ \sum_{x=1}^n p_x p(a(x)|x)_{\rho}\right\},
\end{equation}
where $\{p_x\}$ is a probability distribution given a priori and $p(a(x))|x)_{\rho}$ is the probability of $a(x)$ when measure $x$. This quantity forms a fine-grained uncertainty relation for this set of measurement settings. For the non-commuting observables this quantity is bounded by 1. In \cite{R2014}, the author considered a special fine-grained uncertainty relations of MUBs by letting $p_x = \frac{1}{n},$ for every $x.$ There was obtained an upper bound of $\xi_{\vec{a}}$ for all possible strings $\vec{a}.$ 
Namely, we have the following
\begin{proposition}[\cite{R2014}]
Let $\mathcal{M} = \{M_x: x=1,\cdots,n\}$ be a set of MUBs in a $d$-dimensional Hilbert space. For an arbitrary density matrix $\rho$, we have
{\small\begin{equation}
\frac{1}{n} \sum_{x=1}^n \mathrm{Tr} \left( \left|\phi_x^a \rangle\langle \phi_x^a \right| \rho \right) \leq \frac{1}{d} \left(1+ \frac{d-1}{\sqrt{n}} \right),\;\; \forall a=1,\cdots,d.
\end{equation}}
Therefore, $\xi_{\vec{a}} \leq \frac{1}{d} \left(1+ \frac{d-1}{\sqrt{n}}\right)$ where we have chosen $p_x = \frac{1}{n}.$
\end{proposition}

Using the above proposition we can obtain an alternative violation of the steering inequality defined by steering functional $F$ (see equation \eqref{eq:steerineqMUBs}). 
To end with, let us consider the LHS bound firstly. 
Assume that $\sigma\in\mathcal{L}$. Then
{\small\begin{equation}\label{eq:MUB}
\begin{split}
\langle F,\sigma\rangle & =\sum_{x=1}^{n}\sum_{a=1}^{d}\text{Tr}(\left|\phi_{x}^{a}\left\rangle \right\langle \phi_{x}^{a}\right|\sum_{\lambda}q_{\lambda}p_{\lambda}(a|x)\sigma_{\lambda})\\
& \leq \sum_\lambda q_\lambda \sum_{x=1}^n \left(\sup_a \text{Tr} (\left|\phi_{x}^{a}\left\rangle \right\langle \phi_{x}^{a}\right| \sigma_\lambda)\right) \left(\sum_{a=1}^d p_\lambda (a|x)\right)\\
&  \leq n \sup_a \xi_{\vec{a}} \leq \frac{n}{d} \left(1+ \frac{d-1}{\sqrt{n}} \right).
\end{split}
\end{equation}}
Since the above inequality holds for any $\sigma\in\mathcal{L}$, we get
\begin{equation}
S_{LHS}(F)\leq \frac{n}{d} \left(1+ \frac{d-1}{\sqrt{n}} \right).
\end{equation}
Furthermore, 
$S_Q(F)= n$ by Lemma \ref{qvalue}. Thus, we get the following lower bound:
\begin{equation}V(F) \geq \frac{d\sqrt{n}}{\sqrt{n}+d-1}.\end{equation} Still if the dimension $d$ is an integer power of a prime number, the violation is  lower bounded by $O(\sqrt{d}),$ which coincides with the result of Theorem \ref{thm:MUB}.  Authors of \cite{FGG2013} provided conjecture that the MUBs will give the most uncertain measurement results for the special uncertainty relations considered in the same article. It would be an explanation 
why the steering inequality derived by MUBs provides a higher violation than the random one.

\emph{Unbounded violation: Clifford observables}. Now we will focus on the dichotomic case, where 
there are only two outcomes for each input setting. 
Let us consider operators $A_i\in\mathcal{B}(\mathbb{C}^{2^n})$, $i=1,2,\ldots,2^n$ with the following properties:

\begin{enumerate}
\item[i)] $A_{i}^{\dagger}=A_{i},$
\item[ii)] $A_{i}A_{j}+A_{j}A_{i}=2\delta_{ij}\un_{2^{n}}$, 
\item[iii)] the set $\mathcal{A} = \{A_i, i=1,\cdots,2^n\}$
forms a linear basis of $\mathcal{B}\left(\mathbb{C}^{2^{n}}\right).$
\end{enumerate}
The algebra which is generated by these $A_i$'s is the Clifford algebra. A representation of this algebra can be constructed by tensor products
of Pauli matrices \cite{JW1928}. Now choose arbitrary $n$ operators $A_x, x=1,\ldots,n$ from the set $\mathcal{A}.$
We will consider the following projectors $P_x^a:\,x=1,\ldots,n,\;a=1,2$ where
{\small\begin{equation}
\label{eq:steerineqCliff}
P_x^1 = \frac{1}{2}(\un+ A_x),\quad P_x^2 =  \frac{1}{2} (\un- A_x), \quad x=1,\cdots,n.
\end{equation}}
By the  above projectors, we can define a steering functional $F= \{F_x^a = P_x^a - \frac{\un}{2}: \, x=1,\ldots,n,\; a=1,2\},$ i.e,
{\small\begin{equation}
\label{eq:steerineqCliff1}
F_x^1 = \frac{1}{2} A_x,\quad F_x^2 = - \frac{1}{2} A_x, \quad x=1,\cdots,n.
\end{equation}}
As before, firstly we would like to estimate the quantity  $S_Q(F)$ , direct calculation shows that $S_Q(F)= \frac{n}{2}$  and as before to estimate the bound of $S_{LHS}(F)$  we use the operator norm estimation (see appendix B).
\begin{theorem}\label{thm:dicho} If $F$ is a steering functional defined in \eqref{eq:steerineqCliff1}, then we have
\begin{equation}
V\left(F\right)
\geq \sqrt{\frac{n}{2}}.
\end{equation}
\end{theorem}
The proof of the above theorem  is in appendix B.

There is an alternative way to explain this unbounded violation. By using the notion in \cite{AGT2006}, we can define a traceless operator $F_x$ corresponding to $P_x^a$ as $F_x = P_x^1 - P_x^2 = A_x.$ On the other hand, if we only consider projective measurement, we can define a dichotomic assemblage $\sigma_x = \sigma_x^1 - \sigma_x^2.$ Hence if the LHS model exists, then
\begin{equation}
\sigma_x = \sum_\lambda p_\lambda I(x,\lambda) \sigma_\lambda,
\end{equation}
where $I(x,\lambda) = p(1|x,\lambda) - p(2|x,\lambda) \in [-1,1].$ So we can define a dichotomic steering functional $F^{dicho}$ as:
\begin{equation}\label{eq:dicsteering}
\left|\left\langle F^{dicho},\sigma\right\rangle \right| =\text{Tr}\left(\sum_{x=1}^{n} F_{x} \sigma_{x} \right).
\end{equation}
The quantum and the LHS bound can be similarly defined as before. The following corollary holds:

\begin{corollary}\label{cor:dicsteering}
Let $F^{dicho}$ be the dichotomic steering functional corresponding to the one in Theorem \ref{thm:dicho}, i.e, $F^{dicho} = \{A_x: x=1,\cdots,n \},$ then
\begin{equation}
V\left(F^{dicho}\right) \geq \sqrt{\frac{n}{2}} .
\end{equation}
\end{corollary}

The proof is the same as the proof in the Appendix B, we have
\begin{equation}
S_{LHS} (F^{dicho}) \leq \sup_\lambda\left\Vert \sum_{x=1}^{n}I(x,\lambda) A_{x}\right\Vert _{\infty} \leq \sqrt{2n}.
\end{equation}
For the quantum bound, we use the dichotomic assemblage $\sigma_x = \frac{1}{2^n} A_x.$ Thus $S_Q(F^{dicho}) = n.$

\emph{Conclusions}.
 In this paper, we have provided two steering inequalities with the unbounded violation. 
One is derived from MUBs with violation $O(\sqrt{d})$ in the scheme $(d+1,d,d),$ where $d$ is an integer power of the prime number. We obtain this result using a much simpler method than the operator space theory approach.  Interestingly, this violation is connected to the fine-grained uncertainty relations for MUBs. The question: \textit{Do stronger uncertainty relations exist?} appeared here naturally.   Another is constructed by using the basis of Clifford algebra with violation $O(\sqrt{\frac{n}{2}}) \simeq O\left(\sqrt{\log d}\right)$ in the scheme $(n,2,2^n=d)$. Our result shows an interesting property of quantum steering, since there does not exist a bipartite correlation type Bell inequality with unbounded violation \cite{AGT2006,Pisier2012,Tsirelson}. The mathematical reason for our unbounded violation was explained in a companion paper \cite{HMY2014}, by means of the operator space theory. It shows a different property in quantum steering comparing to Bell nonlocality. The question: \textit{Is there large violation in Bell type correlation in dichotomic case?} seems to be a natural conclusion of this result. After this paper was completed ,  the positive answer for the above question was found \cite{YC}.  Most intriguing and interesting would be to find the fine grained uncertainty relations using anti-commuting observables as a follow up to this work. We hope that the results we obtained will allow a better understanding of correlations that exist in quantum systems.

\emph{Acknowledgments}. We thank Andrzej Grudka, Pawe\l{} Horodecki,  Alexey E. Rastegin and Gniewomir Sarbicki
for valuable discussion. This work was partly supported by TEAM project of FNP, Polish Ministry of Science and Higher Education Grant no. IdP2011 000361, ERC AdG grant QOLAPS, EC grant RAQUEL and a NCBiR-CHIST-ERA Project QUASAR. Z. Yin was partly supported by NSFC under Grant No.11301401.
A. Rutkowski was supported  by a postdoc internship decision number DEC\textendash{} 2012/04/S/ST2/00002, from the Polish
National Science Center.

\section*{Appendix A}
\emph{Proof of Lemma 1.}

To start with:
\begin{eqnarray}
\text{Tr}\left({\displaystyle \sum_{x=1}^{n}}{\displaystyle \sum_{a=1}^{m}F_{x}^{a}\sigma_{x}^{a}}\right) & \leq & \text{Tr}\left({\displaystyle \sum_{x=1}^{n}}{\displaystyle \sum_{a=1}^{m}F_{x}^{a}{\displaystyle \sum_{a'}\sigma_{x}^{a'}}}\right)\nonumber \\
 & = & \text{Tr}\left({\displaystyle \sum_{x=1}^{n}}{\displaystyle \sum_{a=1}^{m}F_{x}^{a}{\displaystyle \rho_{x}}}\right)\\
 & = & {\displaystyle \sum_{x=1}^{n}}{\displaystyle \underbrace{\sum_{a=1}^{m}p_{x}(a|x)}_{=1}=n}\nonumber
\end{eqnarray}
It means
\begin{equation}
S_{Q}\left(F\right) \leq n.
\end{equation}
On the other hand, let us choose the assemblage of the form
$\sigma_{x}^{a}=\frac{1}{d}|\phi_x^a\rangle\langle\phi_x^a|=\frac{1}{d}F_x^a$.
By direct calculations one can obtain $\langle F,\sigma\rangle=n$, what means that
\begin{equation}
S_{Q}\left(F\right) \geq n.
\end{equation}
Comparing these results we get
\begin{equation}
S_{Q}\left(F\right)= n.
\end{equation}

\emph{Proof of Theorem 1.}

Let $\sigma=\{\sigma_x^a\}\in\mathcal{L}$, i.e. $\sigma_x^a=\sum_\lambda q_\lambda p_\lambda(x|a)\sigma_\lambda$. Then
{\small\begin{equation}
\begin{split}
\langle F,\sigma\rangle
&=  \text{Tr}\left(\sum_{x=1}^{n}\sum_{a=1}^{d}M_{x}^{a}\sigma_{x}^{a}\right) \\
&=  \sum_{x=1}^{n}\sum_{a=1}^{d}\text{Tr}(\left|\phi_{x}^{a}\left\rangle \right\langle \phi_{x}^{a}\right|\sum_{\lambda}q_{\lambda}p_{\lambda}(a|x)\sigma_{\lambda})\\
 & =  \sum_{\lambda}q_{\lambda}\sum_{x=1}^{n}\sum_{a=1}^{d}\text{Tr}(\left|\phi_{x}^{a}\left\rangle \right\langle \phi_{x}^{a}\right|p_{\lambda}(a|x)\sigma_{\lambda})\\
 & \leq  \sup_{\lambda} \left\Vert \sum_{x=1}^{n}\sum_{a=1}^{d}\left|\psi_{x,\lambda}^{a}\left\rangle \right\langle \psi_{x,\lambda}^{a}\right|\right\Vert _{\infty},\label{eq:suplambda}
\end{split}
\end{equation}}
where $\left|\psi_{x,\lambda}^{a}\right\rangle =\sqrt{p_{\lambda}(a|x)}\left|\phi_{x}^{a}\right\rangle $. Here the last inequality follows from the duality between
$\ell_1(\Omega; S_1^d)$ and $\ell_{\infty}(\Omega; \mathbb{M}_d).$
For any $\lambda$, let
{\small\begin{equation}G_\lambda=\sum_{x,y=1}^{n}\sum_{a,b=1}^{d}\left\langle \psi_{x,\lambda}^{a}\left|\psi_{y,\lambda}^{b}\right.\right\rangle \left|x\left\rangle \right\langle y\right|\otimes\left|a\left\rangle \right\langle b\right|\in\mathbb{M}_n\otimes\mathbb{M}_d.
\end{equation}}
Using the purification of $\sum_{x=1}^{n}\sum_{a=1}^{d}\left|\psi_{x,\lambda}^{a}\left\rangle \right\langle \psi_{x,\lambda}^{a}\right|$ and its Schmidt decomposition, one can show
\begin{equation}
\left\Vert \sum_{x=1}^{n}\sum_{a=1}^{d}\left|\psi_{x,\lambda}^{a}\left\rangle \right\langle \psi_{x,\lambda}^{a}\right|\right\Vert _{\infty}=\left\Vert G_\lambda \right\Vert _{\infty},
\end{equation}

 Observe that coefficients of $G_\lambda$ are given by the following formula
{\small\begin{equation}
\left|G_{x,y,\lambda}^{a,b}\right|=\left(\delta_{xy}\delta_{ab}+\frac{1-\delta_{xy}}{\sqrt{d}}\right)\sqrt{p_\lambda\left(a|x\right)p_\lambda\left(b|y\right)}.
\end{equation}}
We should estimate the norm of $G_\lambda.$ To do this let us consider new
operator:

{\small\begin{equation}
\tilde{G}_\lambda=\sqrt{d}G_\lambda=\sqrt{d}\sum_{x,y=1}^{n}\sum_{a,b=1}^{d}\left\langle \psi_{x,\lambda}^{a}\left|\psi_{y,\lambda}^{b}\right.\right\rangle \left|x\left\rangle \right\langle y\right|\otimes\left|a\left\rangle \right\langle b\right|,
\end{equation}}
and write it in the block form:
\begin{equation}
\tilde{G}_\lambda=\sum_{x,y=1}^{n}\left|x\left\rangle \right\langle y\right|\otimes\Theta_{x,y,\lambda},
\end{equation}

where blocks $\Theta_{x,y,\lambda}$ are given by
{\small\begin{equation}
\Theta_{x,y,\lambda}=\sqrt{d}{\displaystyle \sum_{a,b=1}^{d}\left\langle \psi_{x,\lambda}^{a}\left|\psi_{y,\lambda}^{b}\right.\right\rangle }\left|a\left\rangle \right\langle b\right|\quad\text{for}\quad x\neq y,
\end{equation}}

Observe that off diagonal blocks are of the form
\begin{equation}
\Theta_{x,y,\lambda}=
|\xi_{x,\lambda}\rangle\langle\xi_{y,\lambda}|
\quad\text{for}\quad x\neq y,
\end{equation}
where $|\xi_{x,\lambda}\rangle=\sum_{a=1}^d\sqrt{p_\lambda(x|a)}|a\rangle$,
while the diagonal blocks
\begin{equation}
\Theta_{x,x,\lambda}=\sqrt{d}{\displaystyle \sum_{a=1}^{d}}p_\lambda\left(a|x\right)\left|a\left\rangle \right\langle a\right|.
\end{equation}
Since $\Vert\xi_{x,\lambda}\Vert=1$ for any $x=1,\ldots,n$, one can define unitary transformations $U_{x,\lambda}$ such that
\begin{equation}
U_{x,\lambda}\Theta_{x,y,\lambda}U_{y,\lambda}^{\dagger}=\left|0\left\rangle \right\langle 0\right|\qquad\text{for}\quad x\neq y.
\end{equation}
Let $U_\lambda=\bigoplus_{x}U_{x,\lambda}.$
Then we get
{\small\begin{equation}
\begin{split}
U_\lambda\tilde{G}_\lambda U^{\dagger}_\lambda&=\sum_{x,y=1}^{n}\left|x\left\rangle \right\langle y\right|\otimes U_{x,\lambda}\Theta_{x,y,\lambda}U_{y,\lambda}^{\dagger}\\
&=
\sum_{x=1}^n|x\rangle\langle x|\otimes U_{x,\lambda}\Theta_{x,x,\lambda}U_{x,\lambda}\\
&+\sum_{x\neq y}|x\rangle\langle y|\otimes |0\rangle\langle 0|\\
&=
\sum_{x=1}^n|x\rangle\langle x|\otimes U_{x,\lambda}\Theta_{x,x,\lambda}U_{x,\lambda}\\
&+ \sum_{x,y}|x\rangle\langle y|\otimes |0\rangle\langle 0| - \sum_x|x\rangle\langle x|\otimes|0\rangle\langle 0|.
\end{split}
\end{equation}}
Now we estimate the norm of $\tilde{G}_\lambda$:

{\small\begin{equation}
\begin{split}
\left\Vert \tilde{G}_\lambda \right\Vert  & =  \left\Vert U_\lambda \tilde{G}_\lambda U^{\dagger}_\lambda \right\Vert \\
& \leq
\left\Vert \sum_{x=1}^{n}\left|x\left\rangle \right\langle x\right|\otimes
\sqrt{d}U_{x,\lambda}\left({\displaystyle \sum_{a=1}^{d}}p_\lambda\left(a|x\right)\left|a\left\rangle \right\langle a\right|\right)U_{x,\lambda}\right\Vert\\
&+ \left\Vert\sum_{x,y=1}^{n}\left|x\left\rangle \right\langle y\right|\otimes\left|0\left\rangle \right\langle 0\right|\right\Vert+\left\Vert\sum_x|x\rangle\langle x|\otimes|0\rangle\langle 0|\right\Vert \\
 & =  \left\Vert \sum_{x=1}^{n}\left|x\left\rangle \right\langle x\right|\otimes\sqrt{d}U_{x,\lambda}\left({\displaystyle \sum_{a=1}^{d}}p_\lambda\left(a|x\right)\left|a\left\rangle \right\langle a\right|\right)U_{x,\lambda}\right\Vert\\
 & + n+1\\
 & =  \sqrt{d}\left\Vert \sum_{x=1}^{n}\left|x\left\rangle \right\langle x\right|\otimes U_{x,\lambda}\left({\displaystyle \sum_{a=1}^{d}}p_\lambda\left(a|x\right)\left|a\left\rangle \right\langle a\right|\right)U_{x,\lambda}\right\Vert\\
 & +  n+1\\
 & \leq   \sqrt{d}\sup_{a,x} p_\lambda\left(a|x\right)+1+n \leq \sqrt{d} + n+1.
\end{split}
\end{equation}}
Comparing it with \eqref{eq:suplambda} we get
\begin{equation}\langle F,\sigma\rangle \leq \dfrac{1}{\sqrt{d}}\sup_\lambda\Vert\tilde{G}_\lambda\Vert\leq 1+\dfrac{n+1}{\sqrt{d}}.\end{equation}
Since it holds for any $\sigma\in\mathcal{L}$, we arrived at
\begin{equation}S_{LHS}(F)\leq 1+\frac{n+1}{\sqrt{d}}.\end{equation}



We finish the proof by using Lemma 1.

\section*{Appendix B}

\emph{Proof of Theorem 2}

Since $P_x^1-P_x^2 = A_x$ and $P_x^1+ P_x^2 = \un,$ then for any $\sigma \in \mathcal{L},$ we have

{\footnotesize\begin{equation}
\begin{split}
& \left|\langle F,\sigma\rangle\right|  =  \left|\langle P, \sigma\rangle - \frac{d n}{2}\right| = \left|Tr \left(\sum_\lambda q_\lambda \sigma_\lambda\sum_{x=1}^{n}\sum_{a=1}^{2}P_{x}^{a}p_\lambda (a|x)\right) -\frac{d n}{2} \right|\\
 & =  \left|Tr\left(\sum_{\lambda}q_\lambda \sigma_\lambda \sum_{x=1}^{n} \left(P_{x}^{1}p_\lambda\left(1|x\right)+P_{x}^{2}\left(1-p_\lambda\left(1|x\right)\right)\right)\right) - \frac{d n}{2}\right|\\
 & =  \left| Tr\left(\sum_\lambda q_\lambda \sigma_\lambda \sum_{x=1}^{n}\left(\left(P_{x}^{1}-P_{x}^{2}\right)p_\lambda\left(1|x\right)+\frac{1}{2}\left(P_{x}^{2}+P_{x}^{2}\right)\right)\right)-\frac{d n}{2}\right|\\
 &= \left|Tr\left(\sum_\lambda q_\lambda \sigma_\lambda \sum_{x=1}^{n}\left(\left(P_{x}^{1}-P_{x}^{2}\right)\left(p_\lambda\left(1|x\right)-\frac{1}{2}\right) +\frac{1}{2}\un\right)\right)-\frac{d n}{2}\right| \\
 & \leq  \sup_\lambda\left\Vert \sum_{x=1}^{n}a_{x,\lambda}A_{x}\right\Vert _{\infty},
\end{split}
\end{equation}}
where $a_{x,\lambda} = p_\lambda\left(1|x\right)-\frac{1}{2} \in [-\frac{1}{2}, \frac{1}{2}].$
In this setting it is convenient to associate real numbers in the
set $\left\{ +1,-1\right\} $ with the binary values $\left\{ 1,2\right\} $ using
the correspondence $1\rightarrow+1$ and $2\rightarrow-1$ as is common
when using discrete Fourier analysis of Boolean functions. Then
\begin{equation}
\begin{split}
\left\Vert \sum_{x=1}^{n}a_{x,\lambda}A_{x}\right\Vert _{\infty}^{2} & = \left\Vert \sum_{x=1}^{n}a_{x,\lambda}A_{x}\sum_{y=1}^{n}a_{y,\lambda}A_{y}\right\Vert\\
& \leq \left\Vert 2\sum_{x=1}^{n}\left|a_{x,\lambda}\right|^{2}\un_{2^{n}}\right\Vert \leq \frac{n}{2}.
\end{split}
\end{equation}
Hence
\begin{align}
S_{LHS}\left(F\right) = \sup_{\sigma\in \mathcal{L}}\left|\langle F,\sigma\rangle\right| \leq \sqrt{\frac{n}{2}}.
\end{align}
In the quantum case, we use the assemblage $\sigma_x^a = \frac{1}{2^n}P_x^a, a=1,2,$ then we have: $S_Q \left(F\right)= \frac{n}{2}.$

\end{document}